\documentclass[twocolumn,showpacs,preprintnumbers,amsmath,amssymb]{revtex4}

\usepackage{graphicx}
\usepackage{dcolumn}
\usepackage{bm}
\usepackage{color}

\newcommand{\ket}[1]{|#1\rangle}
\newcommand{\bra}[1]{\langle#1|}

\begin{document}

\title{Quantum discord and multipartite correlations}
\author{Ma{\l}gorzata Okrasa}
 \email{okrasa@merlin.phys.uni.lodz.pl}
\author{Zbigniew Walczak}
 \email{walczak@merlin.phys.uni.lodz.pl}
\affiliation{%
Department of Theoretical Physics, University of Lodz\\
Pomorska 149/153, 90-236 {\L}\'od\'z, Poland, EU}

\date{\today}

\begin{abstract}
Recently, it was realized that quantum discord   
can be seen as the minimal amount of correlations which are 
lost when some local quantum operations are performed.
Based on this formulation of quantum discord, we provide 
a systematical analysis of quantum and classical correlations 
present in both bipartite and multipartite quantum systems. 
As a natural result of this analysis, 
we introduce a new measure of the overall quantum correlations 
which is lower bounded by quantum discord.
\end{abstract}


\maketitle

\section{Introduction}
In quantum information theory, the problem of characterization 
of correlations present in a quantum state has been intensively 
studied during the last two decades (for review, see 
\cite{Horodeccy09, Guhne09}).
The most significant progress has been made in this subject in
the case of bipartite quantum systems, especially low-dimensional 
ones, which have been studied in the framework of paradigm
based on the entanglement-separability dichotomy introduced by Werner 
\cite{Werner89}. 
In particular, in the framework of this approach it has become clear 
that the correlations present in a quantum state can be classified 
as either classical or quantum, where the latter ones cannot exist 
without the former ones which are identified with entanglement.
However, some results showed that quantum correlations cannot 
be only limited to entanglement, because separable quantum states 
can also have correlations which are responsible for the improvements 
of some quantum tasks that cannot be simulated by classical methods 
\cite{Knill98, Braunstein99, Bennett99, Meyer00, Biham04, Datta05,
Datta07}. 
Therefore, there is a need to study correlations from a perspective 
different than the entanglement-separability paradigm.

The first attempt in this direction was made by Ollivier and Zurek
\cite{Ollivier01} who studied quantum correlations from a measurement 
perspective. 
They considered two natural quantum extensions of the classical 
mutual information and showed that their difference, 
called quantum discord, 
can be used as a measure of the quantumness of correlations 
in bipartite quantum states, including separable ones.
Alternative but closely related attempt in going beyond 
the entanglement-separability paradigm  was made independently 
by Henderson and Vedral \cite{Henderson01} who tried to separate 
classical and quantum correlations in bipartite quantum states.
   
Quantum discord became a subject of intensive study in different
contexts after the recent discovery \cite{Datta08, Lanyon08, Datta09} 
that non-classical correlations other than entanglement can be 
responsible for the quantum computational efficiency of deterministic 
quantum computation with one pure qubit \cite{Knill98}. 
Because evaluation of quantum discord involves optimization procedure,
it was analytically computed only for a few families of two-qubit
states \cite{Luo08, Ali10, Lang10}.  
In these cases, examination of the structure of entanglement and
discord showed that quantum discord is a measure of non-classical 
correlations that may include entanglement however, 
discord is an independent measure. 

Moreover, when Markovian and non-Markovian dynamics of discord 
was analyzed 
\cite{Werlang09, Maziero09, Fanchini10,  Maziero10, Wang10}, 
it was discovered that quantum discord and entanglement 
can behave very differently---in contrast with 
entanglement, in considered cases, 
Markovian evolution can never lead to a sudden death 
of discord, while non-Markovian can lead to its sudden birth.

In the context of complete positivity of reduced quantum dynamics, 
it was discovered that an arbitrary unitary evolution for any system 
and environment is described as a completely positive map on the
system iff system and  environment are initially in a zero-discord 
state \cite{Rodriguez08, Shabani09}.
Furthermore, it was shown that only some zero-discord states can be
locally broadcast \cite{Piani08}. 
Remarkably, it was discovered that a random quantum state
possesses in general strictly positive discord and an arbitrarily 
small perturbation of a zero-discord state will generate
discord---in other words zero-discord states are extremely rare
\cite{Ferraro10}.
Recently, a necessary and sufficient condition for the
existence of non-zero quantum discord was obtained \cite{Dakic10}.
Furthermore, a natural witness for quantum discord for $2 \times N$
states was provided \cite{Bylicka10}.
Moreover, the notion of quantum discord was also extended to
continuous variable systems to study correlations in two-mode 
Gaussian states \cite{Giorda10, Adesso10}.

In this article, we provide a systematical analysis of quantum and
classical correlations present in bipartite quantum systems using an
alternative formulation of quantum discord. 
As a natural result of this analysis, 
we introduce a new measure of the overall bipartite 
quantum correlations and we show that this measure is lower bounded 
by quantum discord.
Finally, we generalize a notion of quantum discord to multipartite 
quantum systems, by invoking quantum relative entropy, 
and then we show that our approach to quantification of correlations
can be naturally extended to multipartite quantum systems.

\section{Quantum discord in bipartite systems}
Let us consider two quantum systems, $A$ and $B$, 
in a state $\rho_{AB}$. 
In quantum information theory, the quantum mutual information 
of a state $\rho_{AB}$,
\begin{equation} 
\label{QMI}
{\cal I}(\rho_{AB}) = S(\rho_{A}) + S(\rho_{B}) - S(\rho_{AB}),
\end{equation}
is regarded as a measure of the total correlations
(classical and quantum) present in a state $\rho_{AB}$, 
where $\rho_{A(B)}$ is the reduced state of the system $A(B)$, 
and $S(\rho) = - \text{Tr}(\rho \log_{2} \rho)$ is the von Neumann
entropy. 
The quantum conditional entropy, 
$S(\rho_{B|A}) = S(\rho_{AB}) - S(\rho_{A})$, 
allows one to rewrite the quantum mutual information 
in the following form 
\begin{equation}
\label{QMI2}
{\cal I}(\rho_{AB}) = S(\rho_{B}) - S(\rho_{B|A}).
\end{equation}  
The fact that the quantum conditional entropy quantifies 
the ignorance about the system $B$ that remains if we make
measurements on the system $A$ allows one to find an alternative 
expression for the quantum conditional entropy, and thereby for 
the quantum mutual information. 

If the von Neumann projective measurement, 
described by a complete set of one-dimensional
orthogonal projectors, $\{\Pi_{i}^{A}\}$, corresponding 
to outcomes $i$, is performed, then the state of the system $B$ 
after the measurement is given by 
$\rho_{B|i} = 
\text{Tr}_{A}[(\Pi_{i}^{A} \otimes I) \rho_{AB}
              (\Pi_{i}^{A} \otimes I)]/p_{i}^{A}$,
where $p_{i}^{A} = \text{Tr}[(\Pi_{i}^{A} \otimes I) \rho_{AB}]$.
The von Neumann entropies $S(\rho_{B|i})$, weighted by probabilities 
$p_{i}^{A}$, lead to the quantum conditional entropy of the system
$B$ given the complete measurement $\{\Pi_{i}^{A}\}$ on the system $A$
\begin{equation}
S_{\{\Pi_{i}^{A}\}}(\rho_{B|A}) = \sum_{i} p_{i}^{A} S(\rho_{B|i}),
\end{equation} 
and thereby the quantum mutual information, induced by 
the von Neumann measurement performed on the system $A$, is defined by
${\cal J}_{\{\Pi_{i}^{A}\}}(\rho_{AB}) = 
S(\rho_{B}) - S_{\{\Pi_{i}^{A}\}}(\rho_{B|A})$.
The measurement independent quantum mutual information 
${\cal J}_{A}(\rho_{AB})$, defined by
\begin{subequations}
\begin{align}
{\cal J}_{A}(\rho_{AB}) & = \sup_{\{\Pi_{i}^{A}\}} 
{\cal J}_{\{\Pi_{i}^{A}\}}(\rho_{AB}) \\
& = S(\rho_{B}) - \inf_{\{\Pi_{i}^{A}\}} \sum_{i} p_{i}^{A} S(\rho_{B|i}),
\end{align}
\end{subequations}
is interpreted as a measure of classical correlations,
${\cal C}_{A}(\rho_{AB}) = {\cal J}_{A}(\rho_{AB})$
\cite{Henderson01, Ollivier01}.
In general case, ${\cal I}(\rho_{AB})$ and ${\cal J}_{A}(\rho_{AB})$ 
may differ and the difference which is interpreted as a measure 
of quantum correlations, 
\begin{subequations}
\label{DA}
\begin{align}
{\cal D}_{A}(\rho_{AB}) 
& = {\cal I}(\rho_{AB}) - {\cal C}_{A}(\rho_{AB}) \\
& = S(\rho_{A}) - S(\rho_{AB}) + 
\inf_{\{\Pi_{i}^{A}\}} \sum_{i} p_{i}^{A} S(\rho_{B|i}),
\end{align}
\end{subequations}
is called quantum discord  \cite{Ollivier01}. 
It is obvious that, in general, the quantum discord 
${\cal D}_{A}(\rho_{AB})$ is not symmetric with respect to 
the systems $A$ and $B$.
However, swapping a role of $A$ and $B$ one can easily get
\begin{subequations}
\begin{align}
{\cal D}_{B}(\rho_{AB}) 
& = {\cal I}(\rho_{AB}) - {\cal C}_{B}(\rho_{AB}) \\
& = S(\rho_{B}) - S(\rho_{AB}) + 
\inf_{\{\Pi_{j}^{B}\}} \sum_{j} p_{j}^{B} S(\rho_{A|j}),
\end{align}
\end{subequations}
where now  the von Neumann projective measurement, 
described by a complete set of one-dimensional
orthogonal projectors, $\{\Pi_{j}^{B}\}$, corresponding 
to outcomes $j$, is performed on the system $B$, 
and the state of the system $A$ after the measurement is given by 
$\rho_{A|j} = 
 \text{Tr}_{B}[(I \otimes \Pi_{j}^{B}) \rho_{AB}
               (I \otimes \Pi_{j}^{B})]/p_{j}^{B}$,
where $p_{j}^{B} = \text{Tr}[(I \otimes \Pi_{j}^{B}) \rho_{AB}]$.

\section{Correlations in bipartite systems}
Recently, it was realized that the quantum discord 
${\cal D}_{A}(\rho_{AB})$ can be expressed alternatively as the
minimal loss of correlations caused by the non-selective von Neumann 
projective measurement performed on the system $A$ \cite{Luo10}
\begin{equation}
{\cal D}_{A}(\rho_{AB}) = \inf_{\{\Pi_{i}^{A}\}} [{\cal I}(\rho_{AB}) - 
 {\cal I} ( {\cal M}_{\{\Pi_{i}^{A}\}}(\rho_{AB}) )],
\end{equation}
where ${\cal M}_{\{\Pi_{i}^{A}\}}(\rho_{AB}) = 
\sum_{i} (\Pi_{i}^{A} \otimes I) \rho_{AB}(\Pi_{i}^{A} \otimes I)$.

In this section, we will explain first why this formulation of quantum
discord is equivalent to its original definition given by equations
(\ref{DA}).
Then, using this formulation of quantum discord we will
investigate correlations present in bipartite systems. 

According to the quantum operations formalism \cite{Kraus83,
Nielsen00} the most general transformation of a quantum state 
$\rho$ can be represented by a linear, completely positive, 
trace-preserving map ${\cal E}$. 
A quantum operation ${\cal E}$ can be written in a form known as 
the operator-sum representation
${\cal E}(\rho) = \sum_{i} E_{i} \rho E_{i}^\dagger$  
where operation elements $\{E_{i}\}$, called the Kraus operators, 
satisfy the completeness relation
$\sum_{i} E_{i}^{\dagger} E_{i} = I$.

Therefore, we see that from the viewpoint of quantum operations 
formalism, the non-selective von Neumann projective measurement
performed on the system $A$ is a local quantum operation, 
${\cal M}_{\{\Pi_{i}^{A}\}}$, with operation elements 
$\{ \Pi_{i}^{A} \otimes I\}$. 

Since the quantum mutual information ${\cal I}(\rho_{AB})$
does not increase under local quantum operations \cite{Nielsen00}, 
therefore the difference 
${\cal I}(\rho_{AB}) - {\cal I}({\cal M}_{\{\Pi_{i}^{A}\}}(\rho_{AB}))$
describes the correlations loss under local quantum operation 
${\cal M}_{\{\Pi_{i}^{A}\}}$.

Let us note that the joint state of systems $A$ and $B$ 
after performing the non-selective von Neumann projective 
measurement on the system $A$ is given by
\begin{equation}
\label{ArhoAB}
{\cal M}_{\{\Pi_{i}^{A}\}}(\rho_{AB}) = 
\sum_{i} p_{i}^{A} \Pi_{i}^{A} \otimes \rho_{B|i}
\end{equation}
whereas the following equations describe
the state of system $A$ and $B$, respectively 
\begin{subequations}
\label{trArhoAB}
\begin{align}
\text{Tr}_{B} [{\cal M}_{\{\Pi_{i}^{A}\}}(\rho_{AB})] & = 
\sum_{i} p_{i}^{A} \Pi_{i}^{A}, \\
\text{Tr}_{A} [{\cal M}_{\{\Pi_{i}^{A}\}}(\rho_{AB})] & =
\sum_{i} p_{i}^{A} \rho_{B|i} =  \rho_{B}.
\end{align} 
\end{subequations}
Using the elementary properties of the von Neumann entropy, 
$S\big(\sum_{i} p_{i}^{A} \Pi_{i}^{A} \otimes \rho_{B|i}\big) =
H(p_{i}^{A}) + \sum_{i} p_{i}^{A} S\big(\rho_{B|i}\big)$ and
$S\big(\sum_{i} p_{i}^{A} \Pi_{i}^{A}\big) = H(p_{i}^{A})$
\cite{Nielsen00}, 
where $H(p_{i}^{A}) = -\sum_{i} p_{i}^{A} \log_{2} p_{i}^{A}$ 
is the Shannon entropy, we can now compute the quantum mutual 
information of ${\cal M}_{\{\Pi_{i}^{A}\}}(\rho_{AB})$
via equations (\ref{QMI}), (\ref{ArhoAB}) and (\ref{trArhoAB})  
\begin{align}
{\cal I} ({\cal M}_{\{\Pi_{i}^{A}\}}(\rho_{AB})) 
& = S(\rho_{B})  - \sum_{i} p_{i}^{A} S\big(\rho_{B|i}\big) \nonumber\\
& = {\cal J}_{\{\Pi_{i}^{A}\}}(\rho_{AB}) \label{21}.
\end{align}
Therefore, the correlations loss under local quantum operation 
${\cal M}_{\{\Pi_{i}^{A}\}}$, 
${\cal I}(\rho_{AB}) - {\cal I}({\cal M}_{\{\Pi_{i}^{A}\}}(\rho_{AB}))$,  
is equal to ${\cal I}(\rho_{AB}) - {\cal J}_{\{\Pi_{i}^{A}\}}(\rho_{AB})$.
Consequently, the minimal loss of correlations caused by local quantum
operation ${\cal M}_{\{\Pi_{i}^{A}\}}$ is given by 
\begin{subequations}
\label{DArhoAB}
\begin{align}
& \inf_{\{\Pi_{i}^{A}\}} [{\cal I}(\rho_{AB}) - 
{\cal I} ({\cal M}_{\{\Pi_{i}^{A}\}}(\rho_{AB}))] 
\label{DArhoABa}\\
& =  {\cal I}(\rho_{AB}) -  
\sup_{\{\Pi_{i}^{A}\}} {\cal J}_{\{\Pi_{i}^{A}\}}(\rho_{AB}) 
\label{DArhoABb}\\
& =  {\cal I}(\rho_{AB}) -  {\cal C}_{A}(\rho_{AB}) 
  = {\cal D}_{A}(\rho_{AB}) 
\label{DArhoABc}.
\end{align}
\end{subequations}
This shows that quantum correlations present in a bipartite state
$\rho_{AB}$, as measured by ${\cal D}_{A}(\rho_{AB})$,
can be seen as the minimal amount of correlations which are lost 
when the non-selective von Neumann projective measurement 
is performed on the system $A$.

Let us note that performing the optimal non-selective von Neumann 
projective measurement 
${\cal M}_{\{\widetilde{\Pi}_{i}^{A}\}}$, for which supremum in
equation (\ref{DArhoABb}) is attained, we leave classical correlations
unaffected, because 
${\cal D}_{A}({\cal M}_{\{\widetilde{\Pi}_{i}^{A}\}}(\rho_{AB})) = 0$
\cite{DattaPhD} which implies via equations (\ref{DArhoAB}) that 
\begin{equation}
\label{24}
{\cal C}_{A}({\cal M}_{\{\widetilde{\Pi}_{i}^{A}\}}(\rho_{AB})) = 
{\cal I} ({\cal M}_{\{\widetilde{\Pi}_{i}^{A}\}}(\rho_{AB})) = 
{\cal C}_{A}(\rho_{AB}).
\end{equation} 
Although the measurement ${\cal M}_{\{\widetilde{\Pi}_{i}^{A}\}}$
causes only the loss of quantum correlations 
in the state $\rho_{AB}$, 
according to classification of bipartite quantum states
\cite{Piani08} the state 
\begin{equation}
\label{ArhoABoptimal}
{\cal M}_{\{\widetilde{\Pi}_{i}^{A}\}}(\rho_{AB}) 
= \sum_{i} \tilde{p}_{i}^{A} \widetilde{\Pi}_{i}^{A} 
\otimes \rho_{B|i}
\end{equation}
can have quantum correlations, which are not captured by 
${\cal D}_{A}({\cal M}_{\{\widetilde{\Pi}_{i}^{A}\}}(\rho_{AB}))$, 
because the states $\rho_{B|i}$ 
do not necessarily commute---according to classification of bipartite 
quantum states \cite{Piani08}, if the states $\rho_{B|i}$ commute, 
then the state (\ref{ArhoABoptimal}) has only classical correlations, 
otherwise the state (\ref{ArhoABoptimal}) has classical and quantum
correlations.

In order to investigate quantum correlations present in the state 
${\cal M}_{\{\widetilde{\Pi}_{i}^{A}\}}(\rho_{AB})$, let us note
that quantum discord ${\cal D}_{B}(\rho_{AB})$ can be expressed 
alternatively as the minimal loss of correlations caused by 
the non-selective von Neumann projective measurement performed 
on the system $B$
\begin{subequations}
\label{DBrhoAB}
\begin{align}
& {\cal D}_{B}(\rho_{AB}) = 
\inf_{\{\Pi_{j}^{B}\}} [{\cal I}(\rho_{AB}) - 
{\cal I}\big({\cal M}_{\{\Pi_{j}^{B}\}}(\rho_{AB})\big)] 
\label{DBrhoABa} \\ 
& = {\cal I}(\rho_{AB}) - \sup_{\{\Pi_{j}^{B}\}} 
{\cal J}_{\{\Pi_{j}^{B}\}}(\rho_{AB}) 
\label{DBrhoABb} \\
& =  {\cal I}(\rho_{AB}) -  {\cal C}_{B}(\rho_{AB}),
\label{DBrhoABc}
\end{align}
\end{subequations}
where ${\cal M}_{\{\Pi_{j}^{B}\}}(\rho_{AB}) = 
\sum_{j} (I \otimes \Pi_{j}^{B}) \rho_{AB}(I \otimes \Pi_{j}^{B})$.

It is clear that when we perform the optimal non-selective von Neumann 
projective measurement ${\cal M}_{\{\widetilde{\Pi}_{j}^{B}\}}$, 
for which supremum in equation (\ref{DBrhoABb}) is attained, 
then the post-measurement joint state is given by 
\begin{align}
\label{BArhoAB}    
{\cal M}_{\{\widetilde{\Pi}_{j}^{B}\}} 
({\cal M}_{\{\widetilde{\Pi}_{i}^{A}\}}(\rho_{AB}))  = 
\sum_{ij} \tilde{p}_{ij}^{AB} \widetilde{\Pi}_{i}^{A}
\otimes \widetilde{\Pi}_{j}^{B},
\end{align}
where $\tilde{p}_{ij}^{AB} = \text{Tr}[(\widetilde{\Pi}_{i}^{A} 
\otimes \widetilde{\Pi}_{j}^{B}) \rho_{AB}]$.  
Let us note that performing this measurement, we leave 
classical correlations unaffected, because 
${\cal D}_{B} ({\cal M}_{\{\widetilde{\Pi}_{j}^{B}\}} 
({\cal M}_{\{\widetilde{\Pi}_{i}^{A}\}}(\rho_{AB}))) = 0$
\cite{DattaPhD} which implies via equations (\ref{DBrhoAB}) that 
\begin{align}
\label{34}
{\cal C}_{B}({\cal M}_{\{\widetilde{\Pi}_{j}^{B}\}} 
({\cal M}_{\{\widetilde{\Pi}_{i}^{A}\}}(\rho_{AB}))) & = 
{\cal I} ({\cal M}_{\{\widetilde{\Pi}_{j}^{B}\}} 
({\cal M}_{\{\widetilde{\Pi}_{i}^{A}\}}(\rho_{AB}))) \nonumber\\ 
& = {\cal C}_{B}({\cal M}_{\{\widetilde{\Pi}_{i}^{A}\}}(\rho_{AB})),
\end{align} 
where equations (\ref{DBrhoAB}) were applied to the state 
${\cal M}_{\{\widetilde{\Pi}_{i}^{A}\}}(\rho_{AB})$ 
instead of $\rho_{AB}$.
The above considerations show that the measurement 
${\cal M}_{\{\widetilde{\Pi}_{j}^{B}\}}$ causes only 
the loss of quantum correlations in the state 
${\cal M}_{\{\widetilde{\Pi}_{i}^{A}\}}(\rho_{AB})$.
According to classification of bipartite quantum states 
\cite{Piani08}, 
the resulting state (\ref{BArhoAB}) has only classical correlations.

Since we have shown that the subsequent optimal
measurements ${\cal M}_{\{\widetilde{\Pi}_{i}^{A}\}}$ and 
${\cal M}_{\{\widetilde{\Pi}_{j}^{B}\}}$ performed on systems $A$ and
$B$, respectively, lead only to the loss of all 
quantum correlations leaving classical correlations
unaffected, and since we know exaclty 
the amount of quantum correlations which are lost 
when the optimal local measurements are performed,  
we can introduce, in a natural way, a new measure of the overall quantum
correlations present in a bipartite state $\rho_{AB}$ which is based 
on quantum discord 
\begin{equation} 
\label{QrhoAB}
Q(\rho_{AB}) = {\cal D}_{A}(\rho_{AB}) + 
{\cal D}_{B}({\cal M}_{\{\widetilde{\Pi}_{i}^{A}\}}(\rho_{AB})).
\end{equation}
As an illustrative simple example, let us consider two
qubits in the state 
$\ket{\psi}_{AB} = (\ket{0} \ket{0} + \ket{1} \ket{+}) / \sqrt{2}$, 
where
$\ket{+} = (\ket{0} + \ket{1}) / \sqrt{2}$.
It can be verified that  
${\cal D}_{A}(\rho_{AB}) = 
2 - \frac{1}{4} [(2+\sqrt{2}) \log_{2}(2+\sqrt{2})
  + (2-\sqrt{2}) \log_{2}(2-\sqrt{2})] \simeq 0.600876$, 
and the optimal measurement 
${\cal M}_{\{\widetilde{\Pi}_{i}^{A}\}}$ 
is described by  
$\widetilde{\Pi}_{0}^{A}  = \ket{0} \bra{0}$ and  
$\widetilde{\Pi}_{1}^{A}  = \ket{1} \bra{1}$.   
Let us note that the post-measurement state 
${\cal M}_{\{\widetilde{\Pi}_{i}^{A}\}}(\rho_{AB}) = 
\frac{1}{2} (\ket{0} \bra{0} \otimes \ket{0} \bra{0} +
\ket{1} \bra{1} \otimes \ket{+} \bra{+})$ 
has quantum correlations.
It can be verified that 
${\cal D}_{B}({\cal M}_{\{\widetilde{\Pi}_{i}^{A}\}}(\rho_{AB})) = 
3 - \frac{1}{2} [(2+\sqrt{2}) \log_{2}(2+\sqrt{2})
  + (2-\sqrt{2}) \log_{2}(2-\sqrt{2})] \simeq 0.201752$, 
and the optimal measurement 
${\cal M}_{\{\widetilde{\Pi}_{j}^{B}\}}$ 
is described by  
$\widetilde{\Pi}_{0}^{B}  = 
(\sin\frac{\pi}{8} \ket{0} + \cos\frac{\pi}{8} \ket{1})
(\sin\frac{\pi}{8} \bra{0} + \cos\frac{\pi}{8} \bra{1})$ and  
$\widetilde{\Pi}_{1}^{B}  = 
(\cos\frac{\pi}{8} \ket{0} - \sin\frac{\pi}{8} \ket{1})
(\cos\frac{\pi}{8} \bra{0} - \sin\frac{\pi}{8} \bra{1})$.
Since the subsequent optimal measurements performed on systems 
$A$ and $B$, respectively lead only to the loss of all 
quantum correlations leaving classical correlations unaffected, 
thus we see that the overall quantum
correlations present in the state $\ket{\psi}_{AB}$ are quantified by
$Q(\rho_{AB}) = {\cal D}_{A}(\rho_{AB}) + 
{\cal D}_{B}({\cal M}_{\{\widetilde{\Pi}_{i}^{A}\}}(\rho_{AB})) 
\simeq 0.802628.$

Let us note that equation (\ref{QrhoAB}) can be rewritten in the
following form via equations (\ref{DArhoABc}), (\ref{DBrhoABc}) 
and (\ref{24})
\begin{equation} 
\label{QrhoAB2}
Q(\rho_{AB}) = {\cal I}(\rho_{AB}) - 
{\cal C}_{B}({\cal M}_{\{\widetilde{\Pi}_{i}^{A}\}}(\rho_{AB})),
\end{equation}
where equation (\ref{DBrhoABc}) was applied to the state 
${\cal M}_{\{\widetilde{\Pi}_{i}^{A}\}}(\rho_{AB})$.
Therefore, we see that the overall bipartite classical 
correlations are given by
\begin{equation} 
\label{CrhoAB}
C(\rho_{AB}) = 
{\cal C}_{B}({\cal M}_{\{\widetilde{\Pi}_{i}^{A}\}}(\rho_{AB})).
\end{equation}

From equation (\ref{QrhoAB}) it follows that in general case 
the quantum discord ${\cal D}_{A}(\rho_{AB})$ 
underestimates the bipartite quantum correlations,
${\cal D}_{A}(\rho_{AB}) \leq Q(\rho_{AB})$.
In other words the quantum discord ${\cal D}_{A}(\rho_{AB})$ 
is a lower bound for the overall quantum
correlations present in a bipartite state $\rho_{AB}$. 
From the other hand, the Henderson--Vedral measure of classical 
correlations, ${\cal C}_{A}(\rho_{AB})$, 
overestimates the bipartite classical correlations because 
\begin{align}
{\cal C}_{A}(\rho_{AB}) & =
{\cal I}(\rho_{AB}) - {\cal D}_{A}(\rho_{AB}) \nonumber\\ 
& \geq {\cal I}(\rho_{AB}) - Q(\rho_{AB}) 
= C(\rho_{AB}),
\end{align} 
which means that ${\cal C}_{A}(\rho_{AB})$ is an upper bound for the
overall classical correlations present in a bipartite state $\rho_{AB}$.
Let us note that $C(\rho_{AB})$ can be rewritten,  
via equations (\ref{CrhoAB}), (\ref{34}) and (\ref{BArhoAB}),
in the form which coincides with the measure of classical correlations
proposed in \cite{Terhal02} 
\begin{align}
\label{CrhoAB2}
C(\rho_{AB}) & = {\cal I} ({\cal M}_{\{\widetilde{\Pi}_{j}^{B}\}} 
({\cal M}_{\{\widetilde{\Pi}_{i}^{A}\}}(\rho_{AB})))\nonumber\\
& = H(\tilde{p}_{i}^{A}) + H(\tilde{p}_{j}^{B}) - 
H(\tilde{p}_{ij}^{AB}) = {\cal I} (\tilde{p}_{ij}^{AB}),
\end{align}
where  ${\cal I} (\tilde{p}_{ij}^{AB})$ is the classical mutual
information for  the joint probability distribution
$\tilde{p}_{ij}^{AB}$.
Taking this into account we can rewrite $Q(\rho_{AB})$, 
via equations (\ref{QrhoAB2}) and (\ref{CrhoAB}), as follows
\begin{align}
\label{QrhoAB3}
Q(\rho_{AB}) & = {\cal I}(\rho_{AB}) - 
                 {\cal I} (\tilde{p}_{ij}^{AB}),
\end{align}
which shows explicitly that the measure of the
overall quantum correlations $Q(\rho_{AB})$ is symmetric with
respect to the systems $A$ and $B$, because both mutual informations
${\cal I}(\rho_{AB})$ and ${\cal I} (\tilde{p}_{ij}^{AB})$ are
symmetric.

Let us note finally that the above results 
shed new light on some recent
developments and help to better understand them. 
Recently, it was numerically verified  that for
two-qubit states with maximally mixed reduced states,
$\rho_{A} = \rho_{B} = \frac{1}{2} I$, we have 
${\cal D}_{A}(\rho_{AB}) = Q(\rho_{AB})$ \cite{Maziero10}. 
In the framework of our approach, this result can be obtained
analytically. It follows directly, via equation (\ref{QrhoAB}), 
from the fact that for these states 
${\cal D}_{B}({\cal M}_{\{\widetilde{\Pi}_{i}^{A}\}}(\rho_{AB})) = 0$,
because the states $\rho_{B|i}$ in equation (\ref{ArhoABoptimal}) 
commute as one can easily check. 
More recently, it has been reported that a zero-discord two-qubit
$X$-state can have quantum correlations \cite{Ficek11}. 
This result can be easily explained in the framework of our approach. 
In particular, from equation (\ref{QrhoAB}) it follows immediately 
that the nullity of quantum discord does not necessarily imply 
the vanishing of quantum correlations.

\section{Correlations in multipartite systems}
In this section, we will show that a notion of quantum discord 
can be extended in a natural way to multipartite quantum systems 
by invoking quantum relative entropy. 
Then, we will find  the overall quantum and classical correlations
present in these systems.

The quantum relative entropy of a state $\rho$ 
with respect to a state $\sigma$ is defined as
$S(\rho||\sigma) =  - S(\rho) - \text{Tr} (\rho \log_{2} \sigma)$.  
The quantum mutual information (\ref{QMI})
is only a special case of quantum relative entropy, 
namely it is the quantum relative entropy of $\rho_{AB}$ with respect
to $\rho_{A} \otimes  \rho_{B}$,  
${\cal I}(\rho_{AB}) = S(\rho_{AB}||\rho_{A} \otimes  \rho_{B})$
(see e.g., \cite{Nielsen00}).
Therefore, in this way we can naturally generalize 
a notion of quantum mutual information to multipartite systems 
and thereby a notion of quantum discord via quantum relative entropy.


Let us consider $m$ quantum systems, $A_{1} \dots A_{m}$, in a state
$\rho_{\bf A}$. 
The quantum mutual information of a state $\rho_{\bf A}$ 
is given by
\begin{align}
\label{MQMI}
{\cal I}(\rho_{\bf A}) & = 
S(\rho_{\bf A}||
\rho_{A_{1}} \otimes  \cdots \otimes \rho_{A_{m}}) \nonumber\\ 
& = \sum_{i} S(\rho_{A_{i}}) - S(\rho_{\bf A}),
\end{align}   
which allows us to define quantum discord for a $m$-partite system.

The quantum conditional entropy, 
$S(\rho_{[A_{k}]|A_{k}}) = 
S(\rho_{\bf A}) - S(\rho_{A_{k}})$, 
allows one to rewrite the quantum mutual information 
in the following form 
\begin{equation}
\label{MQMI2}
{\cal I}(\rho_{\bf A}) 
= \sum_{i \neq k} S(\rho_{A_{i}})  
- S(\rho_{[A_{k}]|A_{k}}),
\end{equation}  
where $[A_{k}]$ stands for 
$A_{1} \dots A_{k-1} A_{k+1} \dots A_{m}$.
The fact that the quantum conditional entropy quantifies 
the ignorance about the systems 
$[A_{k}]$ 
that remains if we make measurements on the system $A_{k}$ 
allows one to find an alternative 
expression for the quantum conditional entropy, and thereby for 
the quantum mutual information.

If the von Neumann projective measurement, $\{\Pi_{i}^{A_{k}}\}$, 
corresponding to outcomes $i$, is performed then 
the post-measurement joint state of the systems 
$[A_{k}]$ is given by 
\begin{equation}
\label{BC|i}
\rho_{[A_{k}]|i} = 
\text{Tr}_{A_{k}}[{\cal P}_{i}^{A_{k}} \rho_{\bf A} 
                  {\cal P}_{i}^{A_{k}}]/p_{i}^{A_{k}}, 
\end{equation}  
where ${\cal P}_{i}^{A_{k}} = (I \otimes \cdots \otimes \Pi_{i}^{A_{k}} 
\otimes  \cdots \otimes I)$ and 
$p_{i}^{A_{k}} = \text{Tr}[{\cal P}_{i}^{A_{k}} \rho_{\bf A}]$.
The von Neumann entropies 
$S(\rho_{[A_{k}]|i})$, 
weighted by probabilities $p_{i}^{A_{k}}$, 
lead to the quantum conditional entropy of the systems
$[A_{k}]$ 
given the complete measurement $\{\Pi_{i}^{A_{k}}\}$ 
on the system $A_{k}$
\begin{equation}
S_{\{\Pi_{i}^{A_{k}}\}}
(\rho_{[A_{k}]|i}) = 
\sum_{i} p_{i}^{A_{k}} 
S(\rho_{[A_{k}]|i}),
\end{equation} 
and thereby the quantum mutual information, induced by 
the von Neumann measurement performed on the system $A_{k}$, 
is defined by
\begin{equation}
{\cal J}_{\{\Pi_{i}^{A_{k}}\}}(\rho_{\bf A}) = 
\sum_{i \neq k} S(\rho_{A_{i}})  
- S_{\{\Pi_{i}^{A_{k}}\}} 
(\rho_{[A_{k}]|A_{k}}).
\end{equation} 
The measurement independent quantum mutual information 
${\cal J}_{A_{k}}(\rho_{\bf A})$ is defined by
\begin{subequations}
\begin{align}
{\cal J}_{A_{k}}(\rho_{\bf A}) & = 
\sup_{\{\Pi_{i}^{A_{k}}\}} 
{\cal J}_{\{\Pi_{i}^{A_{k}}\}}(\rho_{\bf A}) \\
& = \sum_{i \neq k} S(\rho_{A_{i}}) - 
\inf_{\{\Pi_{i}^{A_{k}}\}} 
\sum_{i} p_{i}^{A_{k}} 
S(\rho_{[A_{k}]|i}).
\end{align}
\end{subequations}
Therefore, we define the quantum  discord 
${\cal D}_{A_{k}}(\rho_{\bf A})$
as follows
\begin{subequations}
\label{DArhoABC}
\begin{align}
{\cal D}_{A_{k}}(\rho_{\bf A}) & = 
{\cal I}(\rho_{\bf A}) - 
{\cal J}_{A_{k}}(\rho_{\bf A}) \\
& = S(\rho_{A_{k}}) - S(\rho_{\bf A}) \nonumber\\
& + \inf_{\{\Pi_{i}^{A_{k}}\}} 
\sum_{i} p_{i}^{A_{k}} 
S(\rho_{[A_{k}]|i}).
\end{align}
\end{subequations}
Thus, ${\cal J}_{A_{k}}(\rho_{\bf A})$ 
can be interpreted as a measure of classical correlations
\begin{align}
\label{CArhoABC}
{\cal C}_{A_{k}}(\rho_{\bf A}) & = 
\sum_{i \neq k} S(\rho_{A_{i}})  
 - \inf_{\{\Pi_{i}^{A_{k}}\}} 
\sum_{i} p_{i}^{A_{k}} 
S(\rho_{[A_{k}]|i}),
\end{align}
and consequently
\begin{equation}
\label{DAkrho}
{\cal D}_{A_{k}}(\rho_{\bf A}) = 
{\cal I}(\rho_{\bf A}) - 
{\cal C}_{A_{k}}(\rho_{\bf A}).
\end{equation}
Of course, the quantum discord 
${\cal D}_{A_{k}}(\rho_{\bf A})$  
can be expressed alternatively as the minimal loss of correlations 
caused by the non-selective von Neumann 
projective measurement performed on the system $A_{k}$
\begin{equation}
\label{DAk}
{\cal D}_{A_{k}}(\rho_{\bf A}) = 
\inf_{\{\Pi_{i}^{A_{k}}\}} [{\cal I}(\rho_{\bf A}) - 
{\cal I} ({\cal M}_{\{\Pi_{i}^{A_{k}}\}}(\rho_{\bf A})],
\end{equation}
where 
${\cal M}_{\{\Pi_{i}^{A_{k}}\}} (\rho_{\bf A}) = 
\sum_{i} {\cal P}_{i}^{A_{k}} \rho_{\bf A} 
{\cal P}_{i}^{A_{k}}$.
Obviously, the optimal measurement 
${\cal M}_{\{\widetilde{\Pi}_{i}^{A_{k}}\}}$,
for which infimum in equation (\ref{DAk}) is attained,  
causes only the loss of quantum correlations.

We can now use the above considerations to investigate quantum
correlations present in a state $\rho_{\bf A}$. 
Let us assume that the optimal non-selective von Neumann projective
measurements ${\cal M}_{\{\widetilde{\Pi}_{i}^{A_{1}}\}}, \dots, 
{\cal M}_{\{\widetilde{\Pi}_{i}^{A_{m}}\}}$ leading to the minimal 
loss of quantum correlations are performed subsequently on $m$ quantum
systems $A_{1} \dots A_{m}$. Clearly, the corresponding
post-measurement states are given by 
\begin{subequations}
\begin{align}
& {\cal M}_{\{\widetilde{\Pi}_{i_{1}}^{A_{1}}\}} 
(\rho_{\bf A}), \\
& {\cal M}_{\{\widetilde{\Pi}_{i_{2}}^{A_{2}}\}}
({\cal M}_{\{\widetilde{\Pi}_{i_{1}}^{A_{1}}\}} 
(\rho_{\bf A})), \\
& \vdots \nonumber \\
& {\cal M}_{\{\widetilde{\Pi}_{i_{m}}^{A_{m}}\}} 
(\dots 
({\cal M}_{\{\widetilde{\Pi}_{i_{1}}^{A_{1}}\}} 
(\rho_{\bf A}))).
\end{align} 
\end{subequations}    
Each of these states can have quantum correlations, except the last
one which has only classical correlations.
Therefore, the subsequent measurements lead to the corresponding 
loss of quantum correlations
\begin{subequations}
\begin{align}
& {\cal D}_{A_{1}} (\rho_{\bf A}), \\
& {\cal D}_{A_{2}} ({\cal M}_{\{\widetilde{\Pi}_{i_{1}}^{A_{1}}\}} 
(\rho_{\bf A})), \\
& {\cal D}_{A_{3}} ({\cal M}_{\{\widetilde{\Pi}_{i_{2}}^{A_{2}}\}}
({\cal M}_{\{\widetilde{\Pi}_{i_{1}}^{A_{1}}\}} 
(\rho_{\bf A}))), \\
& \vdots \nonumber \\
& {\cal D}_{A_{m}} ({\cal M}_{\{\widetilde{\Pi}_{i_{m-1}}^{A_{m-1}}\}} 
(\dots 
({\cal M}_{\{\widetilde{\Pi}_{i_{1}}^{A_{1}}\}} 
(\rho_{\bf A})))).
\end{align} 
\end{subequations} 
Therefore, the overall quantum correlations present in a $m$-partite 
quantum state $\rho_{\bf A}$ are measured by
\begin{align}
Q(\rho_{\bf A}) & =  
{\cal D}_{A_{1}} (\rho_{\bf A}) 
+ {\cal D}_{A_{2}} ({\cal M}_{\{\widetilde{\Pi}_{i_{1}}^{A_{1}}\}} 
(\rho_{\bf A})) \nonumber\\
& + {\cal D}_{A_{3}} ({\cal M}_{\{\widetilde{\Pi}_{i_{2}}^{A_{2}}\}}
({\cal M}_{\{\widetilde{\Pi}_{i_{1}}^{A_{1}}\}} 
(\rho_{\bf A}))) \nonumber\\
& + \cdots + 
{\cal D}_{A_{m}} ({\cal M}_{\{\widetilde{\Pi}_{i_{m-1}}^{A_{m-1}}\}} 
(\dots 
({\cal M}_{\{\widetilde{\Pi}_{i_{1}}^{A_{1}}\}} 
(\rho_{\bf A})))),
\end{align} 
which is a multipartite generalization of the measure (\ref{QrhoAB})
introduced in the previous section.
This equation can be rewritten, via equation (\ref{DAkrho}) and 
due to the fact that each measurement remains classical
correlations unaffected, in the following form 
\begin{align}
Q(\rho_{\bf A}) & =  
{\cal I} (\rho_{\bf A}) 
 - {\cal C}_{A_{m}} ({\cal M}_{\{\widetilde{\Pi}_{i_{m-1}}^{A_{m-1}}\}} 
(\dots 
({\cal M}_{\{\widetilde{\Pi}_{i_{1}}^{A_{1}}\}} 
(\rho_{\bf A})))).
\end{align}
Therefore, the overall multipartite classical correlations are given
by
\begin{align}
C(\rho_{\bf A}) & = 
{\cal C}_{A_{m}} ({\cal M}_{\{\widetilde{\Pi}_{i_{m-1}}^{A_{m-1}}\}} 
(\dots 
({\cal M}_{\{\widetilde{\Pi}_{i_{1}}^{A_{1}}\}} 
(\rho_{\bf A})))).
\end{align}
From equations (\ref{DAkrho}) and (\ref{DAk}) it follows that 
$C(\rho_{\bf A})$ can be rewritten as
\begin{align}
C(\rho_{\bf A}) & = {\cal I}
({\cal M}_{\{\widetilde{\Pi}_{i_{m}}^{A_{m}}\}} 
(\dots 
({\cal M}_{\{\widetilde{\Pi}_{i_{1}}^{A_{1}}\}} 
(\rho_{\bf A})))) \nonumber\\
& = {\cal I} (\sum_{i_{1} \dots i_{m}} 
\tilde{p}_{i_{1} \dots i_{m}}^{A_{1} \dots A_{m}} 
\widetilde{\Pi}_{i_{1}}^{A_{1}} \otimes \cdots \otimes 
\widetilde{\Pi}_{i_{m}}^{A_{m}}),
\end{align}
where $\tilde{p}_{i_{1} \dots i_{m}}^{A_{1} \dots A_{m}} = 
\text{Tr} [(\widetilde{\Pi}_{i_{1}}^{A_{1}} \otimes \cdots \otimes 
\widetilde{\Pi}_{i_{m}}^{A_{m}}) \rho_{\bf A}]$.
From equation (\ref{MQMI}) it follows that  
\begin{align}
C(\rho_{\bf A}) & = 
\sum_{k=1}^{m} H(\tilde{p}_{i_{k}}^{A_{k}}) - 
H(\tilde{p}_{i_{1} \dots i_{m}}^{A_{1} \dots A_{m}}) 
 = {\cal I} (\tilde{p}_{i_{1} \dots i_{m}}^{A_{1} \dots A_{m}}),
\end{align}
where ${\cal I} (\tilde{p}_{i_{1} \dots i_{m}}^{A_{1} \dots A_{m}})$
is the classical mutual information for the joint probability
distribution $\tilde{p}_{i_{1} \dots i_{m}}^{A_{1} \dots A_{m}}$.
Therefore, the overall multipartite quantum correlations can 
be rewritten in the following form
\begin{equation}
Q(\rho_{\bf A}) = {\cal I} (\rho_{\bf A}) -
{\cal I} (\tilde{p}_{i_{1} \dots i_{m}}^{A_{1} \dots A_{m}}).
\end{equation}

Let us note that the nullity of the quantum discord 
${\cal D}_{A_{1}} (\rho_{\bf A})$ does not necessarily 
imply the vanishing of quantum correlations present 
in a multipartite state, because the quantum discord 
${\cal D}_{A_{1}} (\rho_{\bf A})$ is a lower bound 
for the overall multipartite quantum correlations. 
Obviously, multipartite counterpart of the Henderson--Vedral measure of 
classical correlations, 
${\cal C}_{A_{1}}(\rho_{\bf A})$, is an upper bound for 
the overall multipartite classical correlations.

\section{Summary}
Using the  alternative formulation of quantum discord, we have
provided a systematical analysis of quantum and classical correlations
present in bipartite quantum systems. 
In particular, we have introduced a new measure of the overall quantum
correlations, and showed that this measure is lower bounded by quantum
discord.
This implies that a zero-discord state can have quantum correlations.
Finally, we have shown that our approach to quantification of correlations
can be naturally extended to multipartite quantum systems.

\section*{Acknowledgments}
This work was supported by the University of Lodz Grant, 
the Polish Ministry of Science and Higher Education 
Grant No. N N202 103738, and the Polish Research Network 
{\em Laboratory of Physical Foundations of Information Processing}. 
M.~Okrasa acknowledges the support from the European Union under 
the Human Capital Programme -- Measure 8.2.1 and the IROP -- Measure 2.6.


\end{document}